\documentclass{emulateapj}
\usepackage{lscape}
\begin{document}

\title{Optical Spectroscopy of a Flare on Barnard's Star}

\author{Diane B. Paulson\altaffilmark{1} }
\affil{NASA GSFC, Code 693, Geenbelt MD 20771}
\email{dpaulson@lepvax.gsfc.nasa.gov}

\author{Joel C. Allred}
\affil{Physics Department, Drexel University, Philadelphia PA 19104}

\author{Ryan B. Anderson}
\affil{Department of Astronomy, University of Michigan, Ann Arbor MI 48109}
\altaffiltext{1}{A National Research Council Postdoctoral Fellow working at
NASA's Goddard Space Flight Center.} 

\author{Suzanne L. Hawley}
\affil{Department of Astronomy, University of Washington, Seattle WA 98195}

\author{William D. Cochran}
\affil{McDonald Observatory, University of Texas, Austin TX 78712}

\and 

\author{Sylvana Yelda}
\affil{Department of Astronomy, University of Michigan, Ann Arbor MI 48109}

\begin{abstract}
We present optical spectra of a flare on Barnard's star. 
Several photospheric as well as chromospheric
species were enhanced by the flare heating. An analysis of
the Balmer lines shows that their shapes are best explained 
by Stark broadening rather than 
chromospheric mass motions. We estimate the  temperature of the flaring 
region in the lower atmosphere to be $\ge$8000~K and the electron density to be 
$\sim$10$^{14}$~cm~$^{-3}$, 
similar to values observed in other dM flares.
Because Barnard's star is considered to be one of our oldest neighbors, 
a flare of this magnitude is probably quite rare. 

\end{abstract}


\section{Introduction}
Barnard's star (Gliese 699) is a well-studied, nearby (d=1.8 pc, 
\nocite{HuScSt99}H\"unsch et al. 1999, \nocite{Gi96} Giampapa et al. 1996) 
M dwarf (\nocite{MaMiPe00}Marino et al. 
2000, \nocite{HuScSt99}H\"unsch et al. 1999). 
\citet{CiMa04} and \citet{ReHaGi95} classify it as an M4 dwarf, and \citet{Gi96}
provide a mass estimate of 0.144~M$_\odot$.
\citet{DaDR04} measure a bolometric luminosity of
(3.46$\pm$0.17)x10$^{-3}$L$_{\odot}$,  and infer a radius=0.200$\pm$0.008~
R$_{\odot}$, and effective temperature=3134$\pm$102 K.

Barnard's star is considered to be a very old neighbor. \citet{Gi96} find
that Barnard's star lies just below the main sequence, and may be a
Population II subdwarf. 
It also has a very low quiescent X-ray luminosity 
(e.g.,  log($L_{\rm X}$)=26, H\"unsch et al. 1999; 
26.1, \nocite{VaCaFa81}Vaiana et al. 1981), indicating only low level 
magnetic activity.
\citet{BeMACh99} infer a rotation period of 130~days
from interferometric photometry. The slow rotation is another indication 
of advanced age.
Low-level variability on Barnard's star has been
observed (e.g. \nocite{KuEnRo03}K{\"u}rster et al. 2003, 
\nocite{BeMACh99}Benedict et al. 1999), as is common for M dwarfs
\citep[e.g.][]{MaMiPe00}.
Additionally, very young, active M dwarfs typically show Balmer line emission 
during quiesence. The Balmer lines are not present (either in absorption 
or in emission) during quiescence \citep{StHa86, HeLa87} in Barnard's star. 

Marino et al. (2000) noted that their measurement 
(log($L_{\rm X}$)=25) of Barnard's star was taken ``in flare".
Additionally, \citet{RoCrGi90} may have caught 
Barnard's star during a low-level flare, as they detect a slight emission
in H$\alpha$. In this paper, we present a census of enhanced features, 
including the Balmer series, during a flare event that we observed
which is significantly stronger 
than the one in \citet{RoCrGi90}. The observation of flare events on 
old M dwarfs is difficult owing to their infrequency. As such, most 
flares are caught only as snapshots and do not provide complete time 
coverage of the event. However, it is 
interesting to describe the isolated events that are observed and to attempt to 
place them in perspective compared to flares on well-studied flare stars, 
such as AD Leo \citep{HaAlJK03}.
 
\section{Observations}
The echelle spectra were obtained at McDonald Observatory's 2.7~m 
Harlan J. Smith telescope on July 17, 1998
during planned observations for the McDonald Observatory Planet Search
\citep{CoHa94}. 
When the star began flaring, two additional spectra were taken.
The cross-dispersed coude echelle spectrograph was used along with 
the TK3 detector.
The spectral
coverage is almost complete from 3600 \AA\ to 10800 \AA.
Spectra are taken with resolution of $\sim$60,000 and
Considering the 
spectral type of Barnard's star and thus the low S/N in the bluest orders, 
we do not consider any spectral features 
below 3700 \AA\ and we only include lines below $\sim$3800 \AA\ which are 
obviously in emission and easily identified. There are interorder 
gaps in the spectra  redward of 
$\sim$ 5800\AA\. We note in the following sections when important
lines fall in these gaps.

\begin{figure}
\plotone{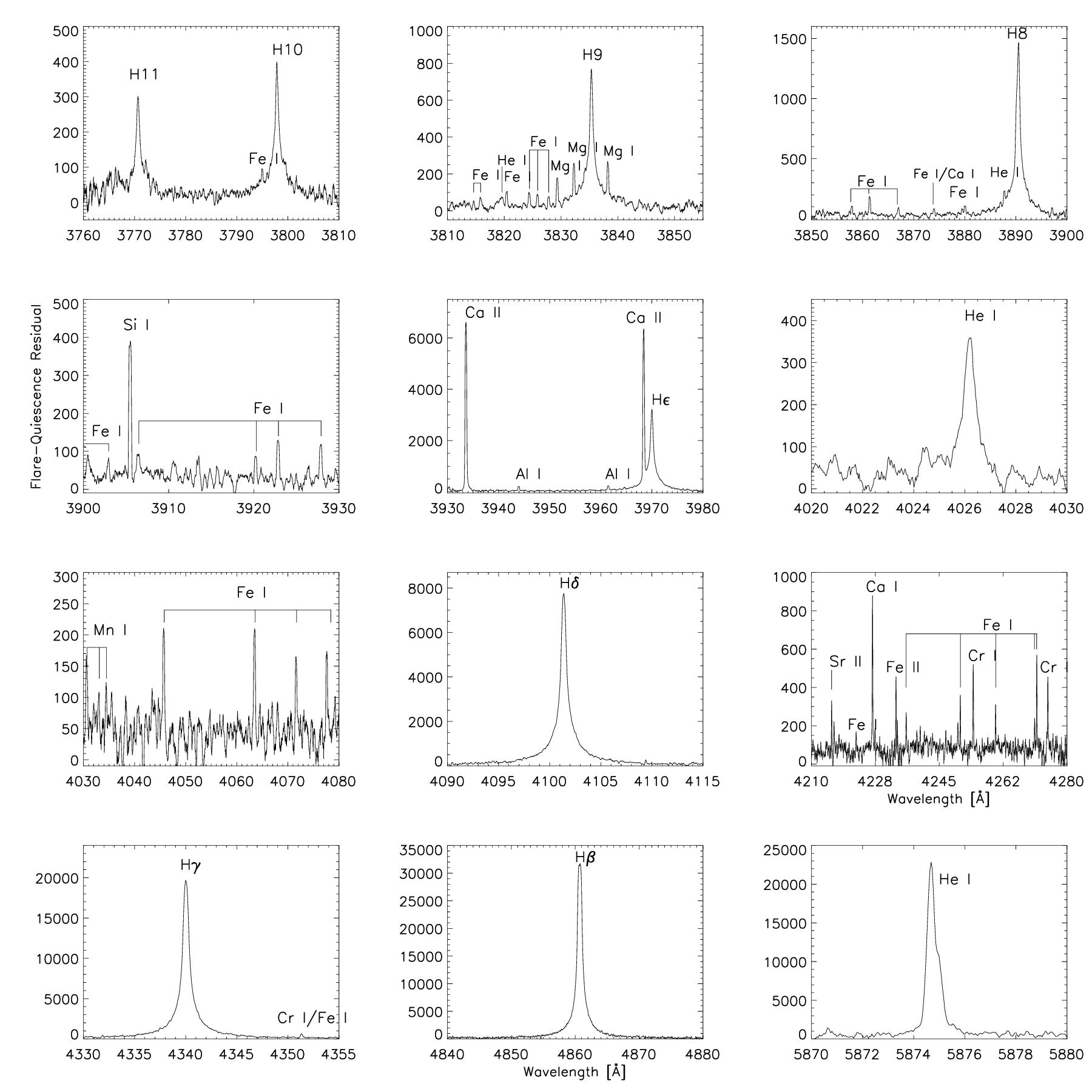}
\caption{Examples of observed spectral regions with interesting
lines. The quiescent spectrum has been subtracted. Spectra have also been
smoothed to a resolution of R=20,000, though the
original data are used in the caluclations described in the text.}
\end{figure}

All data were 
reduced with standard IRAF\footnote{Standard IRAF footnote here.} 
echelle packages. The wavelength scale was derived using the spectrum
of a ThAr calibration lamp. Spectral orders with clearly defined continuua were
fit with 5th order polynomials. For orders which do 
not have clearly defined continuua, we fit 3rd order 
polynomials to define a pseudo-continuum in regions near important lines 
(e.g. in the order containing 
Ca~II~H~\&~K, 
the regions just outside the H~\&~K lines and in between the 
lines were used). Absolute flux calibration of the spectra was not attempted.

The first spectrum taken at 05:32:09.04 UT is in quiescence. 
The following spectrum (F1), taken immediately after the first at 
06:06:09.42 UT,
included the flare maximum. The third spectrum (F2), taken at 06:39:48.58 UT
showed much less flare enhancement, indicating that the flare
decayed rapidly. The quiescent spectrum and F1 were taken with the I$_{2}$
cell in place. Molecular I$_2$ lines litter the spectrum between 5000 and 
6200\AA. These lines subtract out for the most part when differencing these
spectra (for line identification) but as a precaution, the lines we identify 
in $\S$3. are those which are unambiguous (i.e. $\ge$4$\sigma$). Because
F2 was taken without the cell in place, the difference of F2 and quiescent 
spectra contain the I$_{2}$ lines. For these reasons, we have not included 
the lines detected in F2 in $\S$3 except for the Balmer lines which are 
sufficiently large for unambiguous detection, as described in $\S$3.2.1.

\section{Results}
\subsection{Continuum Enhancement}
The continuum is enhanced during the flare, but this enhancement is difficult 
to measure in absolute terms from our echelle observations. 
We assume the relative calibrations as described in $\S$3.2.1 are 
reasonable, and we thus provide
continuum increases for various wavelength intervals in Table 1. 
The flare clearly shows a strong blue color as is typically seen in stellar 
flares \citep{HaPe91}.
A lower limit to the temperature of the flare can be estimated assuming the 
continuum peaks at our bluest wavelength observed.
This gives an estimate of a  blackbody temperature $\ge$ 8000 K. 
This is similar to the temperature derived during an
unusually large flare on the K dwarf LQ~Hya \citep{MoSaCo99} and 
to temperatures derived for other flares on M dwarfs (e.g., Hawley et al 2003).

\begin{figure}
\plotone{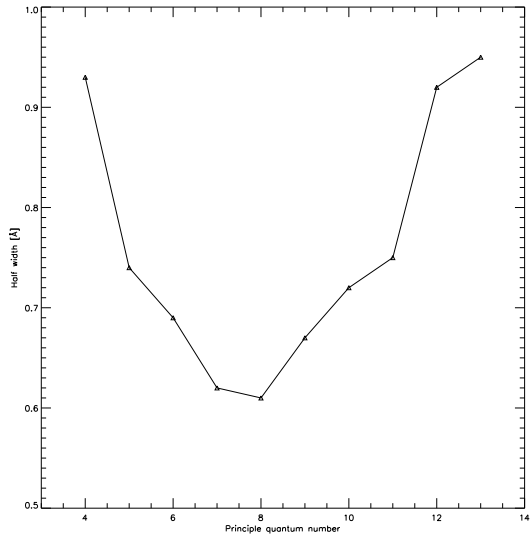}
\caption{The half width of the Balmer lines.}
\end{figure}

\subsection{Line Features}
Table 2 lists the dominant lines filled-in or in emission during the
flare which are included in the wavelength span of our data and which do not
fall in interorder gaps. For example,
only the Na~I~D1 line (5896\AA) is available, because the D2 line
falls in an interorder gap. We adopt excitation potentials and $gf$ values
from \citet{KuBe95}.

\subsubsection{Balmer Series}
The Balmer lines are not present in quiescent spectra of Barnard's star, but are strongly in emission during the flare event. 
The Balmer lines from H$\beta$ to H11 are 
shown in Figure 1, along with several other spectral regions of interest, for
the first flare spectrum with the quiet spectrum subtracted.
Our spectra contain the Balmer series up to H13 with the sole
exception of H$\alpha$ which lies in one of the red interorder gaps. 
The Balmer lines are
significantly broadened, as also noted during several other stellar and solar flares 
\citep[e.g.,][]{HaPe91,Sv72, GAJeDo02}.
As described below, while the H and He lines are broadened, other chromospheric lines such as Ca II H\&K are not broadened during the flare event. This is 
evidence for Stark broadening. The measured Balmer
decrements (relative to H$\gamma$) are listed in Table 3 for both 
F1 and F2. We measured the EWs relative to the normalized continuum. We then 
assigned a flux to the continuum by using a flux calibrated spectrum of 
AD Leo. Multipling the EW by this flux value gives our final Balmer decrement.
However, we note that the spectral type of AD Leo is M3 whereas Barnard's star is
M4. Thus, our decrement measurements are only approximations considering 
the slightly mismatched continuum flux levels.
As expected, the decrement decreases with increased Balmer number
(Table 3) and the magnitude of the half width reverses at H8 (Figure 2). 
The trend in the decrement appears to be 
slightly steeper than the flares on AD~Leo 
\citep[][ and references therein]{HaPe91}. 

\begin{figure}
\plotone{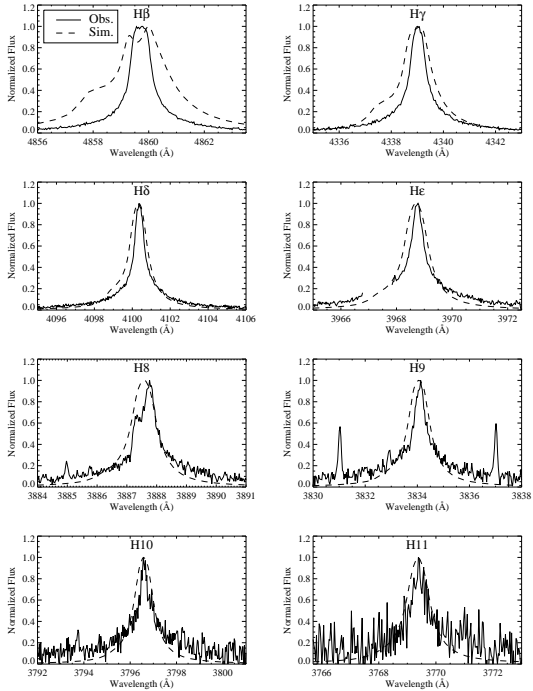}
\caption{Observed Balmer line profiles compared to predictions obtained from
the
F10 flare simulation of Allred et al. (2006).  Observed and simulated line
profiles are indicated by solid and dashed lines respectively.  The line
profiles have been normalized to the peak intensity.}
\end{figure}

\citet{WoScGi84} find that Stark broadening causes $\sim$3
times the FWHM broadening in H9 than in H$\gamma$.  But this is not the
case for our spectra, where
the FWHM (Table 3) of H$\gamma$ is 0.77\AA\ and in H9 is 0.65\AA.
There are at least two possible reasons for this disagreement.
First, there are NLTE effects in the atmosphere which
are not taken into account when applying a strict Stark profile. At
each wavelength, we see a combination of profiles from different layers which
have different electron density and temperature. Because the flux
in the core gets trapped from the optically thick atmosphere and the
flux in the wings is able to escape, the resulting profile is
broader than predicted for the
lower order Balmer lines where the NLTE effects are most important.
Second, our integration time is quite long so the impulsive phase
broadening probably occurred only during a short part of the
exposure.

We can provide estimates of electron densities during the flare.
According to \citet{DrUl80}, our Balmer decrement is 
shallow indicating that the electron densities (n$_e$) 
are $\ge$10$^{13}$~cm$^{-3}$. 
An additional constaint on n$_e$ can be placed using the Inglis-Teller 
relation for Stark broadening.
The highest resolved line is H13 and thus the upper limit on 
n$_e$ is 1.5~x~10$^{14}$~cm$^{-3}$ \citep[e.g.,][]{KuMa70}, an
order of magnitude below flares on AD Leo
\citep{HaPe91} and on YZ Cmi \citep{WoScGi84}. 

\begin{figure}
\plotone{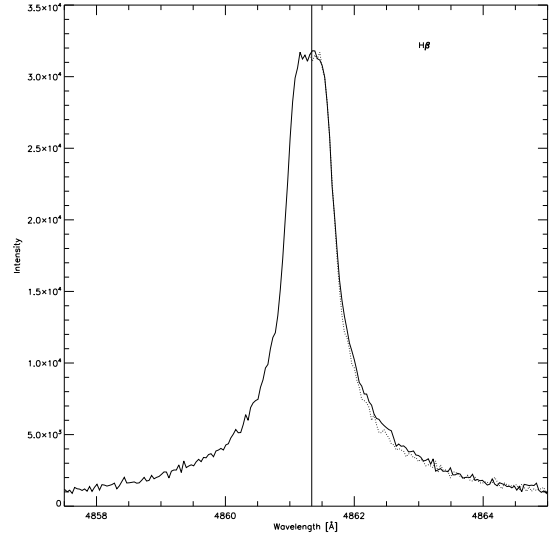}
\caption{The H$\beta$ line with the blue wing
transcribed onto the red wing. There is very slight evidence for asymmetry.}
\end{figure}

A better approach is
to carry out a detailed model of the line profiles using radiative 
hydrodynamical models. In Figure 3 we compare the observed Balmer lines 
during the flare to 
simulated line profiles obtained from a radiative hydrodynamic model of 
flares on M dwarf stars \citep{Allred05}. The simulated line profiles 
were calculated using the radiative transfer code MULTI \citep{Carlsson86}
with a 13 level plus continuum model hydrogen atom. Temperature and 
electron density stratifications for a flaring M dwarf atmosphere were 
taken at numerous times during the F10 dynamical flare simulation reported 
in \citet{Allred05}. Line profiles were computed at each time and the 
results were co-added to produce an average line profile over the duration 
of the simulation. The F10 simulation corresponds to a moderately sized 
flare with an average electron density in the region of Balmer formation 
of $\sim 1.3 \times 10^{13}$~cm$^{-3}$, and is therefore well suited for 
comparison 
to this flare. The predicted line profiles are significantly broader than 
observed for the low order Balmer transitions.  This is likely due to the 
assumption of complete redistribution in the dynamical computation.  The 
large optical depth in the line cores of the lower order lines causes 
emission to be redistributed into the wings and results in broader line 
profiles. The higher order lines, where the effects of partial 
redistribution are less important, more closely match the observations. 
The principal broadening mechanism at these temperatures and densities in 
the simulation is found to be Stark broadening (see discussion in Allred 
et al. 2006).

Asymmetry in the Balmer lines has been reported for several stellar as well as 
solar flares, especially in the
H$\alpha$ and H$\beta$ lines (e.g. \nocite{FuSc04}Fuhrmeister \& Schmitt 2004, 
\nocite{Eason92}Eason et al. 1992, Johns-Krull et al. 1997,
Schmieder et al. 1987\nocite{Schmieder1987}, W\"ulser 1987\nocite{Wulser87},
Canfield et al. 1990\nocite{Canfield1990}). 
These are also seen in the UV lines in 
AD~Leo \citep{HaAlJK03}. The asymmetric nature of chromospheric lines
is attributed to mass motions in the chromosphere- 
the Neupert effect (evaporation) or condensation. The Neupert effect is seen in $\sim$80\%
of large solar flares \citep{MTFiLi99} as well as in some stellar flares
\citep{HaFiSi95, guedel96}. Unfortunately, by 
integrating over 30 minutes, sharp velocity features will likely have
been washed out and thus we are unable to comment on mass motions,
evaporation and velocity fields.  
As an example, Figure 4 shows an expanded version of the 
H$\beta$ line in the F1 spectrum with the quiet spectrum subtracted. The dotted line is
the blue half of the feature transposed onto the red side of the 
line. There is a very slight enhancement of the red wing with respect to the 
blue wing. This asymmetry is not seen in the instrumental profile as measured
by the ThAr calibration lamp taken during the night. The lack of substantial 
asymmetry in our spectra is not surprising;
the asymmetric profile shape should be smeared out
during the course of the integration. The slight asymmetry of H$\beta$ may 
be caused by condensation, but the data are not precise enough for any 
definitive conclusion.

\begin{figure}
\plotone{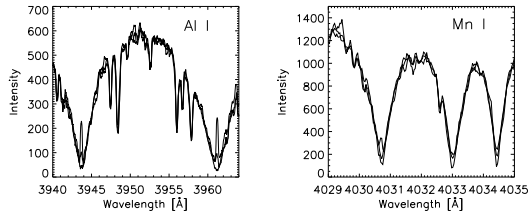}
\caption{The Al~I doublet and Mn~I triplet. Both of the Al~I lines have a prounounced emission core during the first exposure of the flare while only one
of the Mn~I lines shows an emission core though the other members of the
multiplet appear
filled-in. Spectra have been smoothed to a resolution of 20,000, as in Figure
1.}
\end{figure}

\subsubsection{Other Chromospheric Lines}
Several broad He~I lines are in emission during the flare including 
pronounced emission in the 5876\AA\ line. \citet{Zi88} note 
that the He I 5876\AA\ line is in emission only for very strong solar flares 
and is in absorption for medium and small flares. 
In our spectra, the
emission in the He~I lines (including 5876\AA) is only present in F1
and is unmeasurable in F2. The He~I lines are broad lines and the 
contamination of the narrow I$_2$ lines in this region does not affect the 
detection of the broad He feature.
Because we see the He feature in F1 and not F2, the flare was probably quite 
energetic but the impulsive heating was short lived ($<$ 30 min).
The bottom right panel of Figure 1 
shows an example of the fine structure that is present in the He~I lines.  

Additionally, Garcia-Alvarez et al. (2001) discuss the emission
from a nearby Mn~I line (at 4030\AA) possibly confusing the detection of the 
He~I~4026\AA\ line
during times of increased activity. The He~I~4026\AA\ is clearly seen 
during our event (Figure 1), despite line core emission in the
nearby Mn~I line at 4030\AA\ (Figure 5). As discussed in $\S$3.2.3, the Mn~I
4030\AA\ emission core is shallow enough and our spectral resolution is 
sufficiently high to resolve these two lines. Thus the detection of this line is
not compromised by the Mn~I line.

The He~II 4686\AA\ line is not commonly observed in UV~Ceti-type stars
but is present in our spectrum F1. \citet{AbAlAv98} also detected it in spectra
of a flare on EV Lac and noted that it behaved in the same way as the He~II
1640\AA\ line in AD Leo \citep{ByGa88}. \citet{ZiHi85}
discuss the formation of this line in solar flares and conlcude that
it is only formed in very deep regions of the chromosphere during the most
intense flares.

Only one of the Ca~II~IR triplet lines (8662\AA) is available due
to interorder gaps. A strong Fe~I line is
blended with this line but, again our spectral resolution is sufficient so
that it  does not confuse the detection of
filling-in of the Ca~II line. The Fe~I line does not show 
enhancement of any kind during the flare. While the Ca line is 
filled-in, there
is no evidence for an emission core in the 8662\AA\ Ca~II line, though this is
not too surprising as it is the weakest of the triplet \citep{PeCo81}. 

\subsubsection{Enhanced Photospheric Lines}
Several strong lines of neutral metals are seen to have emission
cores during the flare, while others only show a filling-in of the core. 
Typically, the lines that have emission cores, as opposed to just 
filling-in, are very strong lines blueward of 
4000\AA. We suggest that the filling-in effect is caused by the same heating 
as that which produces emission cores but is the limiting case in a 
$relatively$ weak line. 
As an example, Figure 5 shows the pronounced emission in the cores 
of the 3944 \& 3961\AA\ Al~I lines. In quiescence, only the strong 
absorption line is present. The emission seen is only in F1
and by F2 the core has died down significantly, though it is still present. 
Several of the blueward Fe~I lines also show this central core emission. 
This is not often seen in dM flare spectra, 
but have been noted in various solar flares 
\citep[e.g.,][]{CoMa69,JK97} and in selected cases of stellar 
flares \citep{AcFaSa82, HaPe91}. 
Johns-Krull et al. suggest that the 
optical depth in the flaring plasma is large enough to produce optically thin 
emission lines in the cores of strong photospheric lines. Thus, for a weak 
line, it would appear as a filling-in, whereas in a resolved, strong line, 
it would appear as an emission core.

Comparing this flare with that on AD Leo, our spectra show
photospheric line emission in all of the lines listed in Table 2 of 
\citet{HaPe91}, excepting the unidentified lines at 3856.0, 4078.7 and 
4416.1\AA\ and the Fe I line at 4358.51\AA.
Owing to the fact that our spectra are of higher spectral resolution and
higher S/N, we see several lines not identified by Hawley \& Pettersen. 
Comparing our list of identifications with that of \citet{JK97}, who
had similar S/N and resolving power, we note
that redward of $\sim$4900\AA, we only detect enhancement in about half of
the neutral metals they list as enhanced during a 1993 solar flare, whereas 
we detect almost all of the lines listed in their Table 7 blueward of 
4900\AA. The reason for this is likely that the flare on Barnard's
star was more energetic (and thus bluer) than the solar flare or bluer relative
to the stellar photosphere.

It is important to understand the cause of the heating during a flare.
During stellar flares, strong chromospheric lines such as Mg~II~h\&k and
Ca~II~H\&K provide a significant source of radiation. This radiation 
could provide a source of optical pumping for other lines.
The 4030\AA\ Mn~I line has a central core emission similar to the Al~I
lines, while the other Mn~I lines at 4033 and 4035\AA\ only show filling-in 
(Figure 5). \citet{DoOoKe92} suggest that pumping by the Mg~II~k line 
could cause the Mn~I 4030\AA\ line to appear in emission. Because the other 
members of
the Mn~I triplet are also filled-in, optical pumping is unlikely to be the 
cause of this emission. 
Additionally, \citet{He45}  and \citet{Wi74} find that selective 
emission in 
the Fe 43 multiplet (4005.23, 4045.82, 4063.57, 4071.74\AA) \citep{Moore45} may be explained by
optical pumping by the 
Ca~II~H line. We observe filling-in of 6 of the 7 multiplet 43 members. The
3969\AA\ line is completely blended with the H$\epsilon$ line and thus is 
unmeasurable. Our observations do not support the optical pumping mechanism 
to produce the emission in these features. Instead they are probably caused
by the excitation/heating of the upper photosphere and provides further evidence
for in situ heating of the chromosphere.

\section{Summary}
Because Barnard's star is an old M dwarf, strong flares are probably
uncommon. Fortuitously, a flare was observed during an unrelated science
program with a high resolution spectrograph. The flare produced deep 
chromospheric heating resulting in strong blue continuum emission 
and significant Stark broadening in the Balmer emission lines. We determine
a lower limit of 8000K for the blackbody temperature of the flaring region. 
In addition, the upper photosphere was heated sufficiently 
to cause emission in the cores of strong neutral metal lines, which has been
previously observed during solar flares and less frequently in stellar flares.
These data should provide good constaints on the heating of the lower atmosphere
in detailed models of stellar flares.  
\acknowledgments
We thank the referee for providing helpful suggestions for clarification of 
this manuscript.
We thank C. Cowley for useful discussions in the preparation of
this manuscript. SLH and JLA are supported by NSF grant AST-0205875 and
HST grants GO-8613 and AR-10312.
RBA wishes to thank the UROP program for providing him this research 
opportunity.


\clearpage 
\begin{deluxetable}{lc}
\tabletypesize{\scriptsize}
\tablecaption{Continuum Enhancement}
\tablewidth{0pt}
\tablehead{\colhead{wavelength range (\AA)} &
\colhead{\% change during flare maximum}}
\startdata
4170 - 4200	&	4.0$\pm$0.5\\	
4020 - 4080	&	4.8$\pm$0.5\\
3930 - 3990	&	5.2$\pm$1.0\\
3840 - 3900	&	6.5$\pm$1.0\\
3760 - 3810	&	12.6$\pm$3.0\\
3700 - 3730	&       24.0$\pm$5.0\\
3630 - 3670	&       26.3$\pm$5.0\\
\enddata
\end{deluxetable}

\begin{deluxetable}{llcc}
\tabletypesize{\scriptsize}
\tablecaption{Enhanced lines during flare\tablenotemark{1}}
\tablewidth{0pt}
\tablehead{
\colhead{Species} & \colhead{Wavelength (\AA)} & \colhead{$\chi$ (eV)} &
        \colhead{log$gf$}}
\startdata
H13  &  3734.37         &      10.200  &       -1.874   \\
Fe I &  3734.86         &       0.859  &        0.317   \\
H12  &  3750.15         &      10.200  &       -1.764   \\
H11  &  3770.63         &      10.200  &       -1.644   \\
Fe I &  3772.23         &       3.047  &       -2.459   \\
Fe I &  3795.00         &       0.990  &       -0.760   \\
H10  &  3797.90         &      10.200  &       -1.511   \\
Fe I &  3814.52         &       1.011  &       -2.389   \\
Fe I &  3815.84         &       1.485  &        0.298   \\
He I &  3819.613        &       20.96  &       -1.794   \\
He I &  3819.614        &       20.96  &       -1.315   \\
Fe I &  3820.43         &       0.859  &        0.119   \\
Fe I &  3824.44         &       0.000  &       -1.362   \\
Fe I &  3925.20         &       3.292  &       -1.403   \\
Fe I &  3825.88         &       0.915  &       -0.037   \\
Fe I &  3827.83         &       1.557  &        0.062   \\
Mg I &  3829.35         &       2.709  &       -0.207   \\
Mg I &  3832.30         &       2.712  &        0.146   \\
Fe I &  3833.31         &       2.559  &       -1.031   \\
H9   &  3835.39         &      10.200  &       -1.362   \\
Mg I &  3838.290        &       2.717  &       -1.506   \\
Mg I &  3838.292        &       2.717  &        0.415   \\
Mg I &  3838.295        &       2.717  &       -0.333   \\
Fe I &  3849.97         &       1.011  &       -0.871   \\
Fe I &  3850.82         &       0.990  &       -1.734   \\
Fe I &  3852.57         &       2.176  &       -1.236   \\
Fe I &  3856.37         &       0.052  &       -1.286   \\
Fe I &  3859.91         &       0.000  &       -0.710   \\
Fe I &  3865.52         &       1.011  &       -0.982   \\
Fe I/Ca I & 3872.50/3872.54 &   0.990/2.523 & -0.928/-1.070\\
Fe I &  3878.57         &       0.087  &       -1.350   \\
Fe I &  3886.28         &       0.0516  &       -1.075   \\
He I &  3888.646        &      19.820  &       -1.190   \\
He I &  3888.649        &      19.820  &       -0.969   \\
H8   &  3889.05         &      10.200  &       -1.192   \\
Fe I &  3895.66         &       0.110  &       -1.670   \\
Fe I &  3899.71         &       0.087  &       -1.531   \\
Fe I/Cr I &3902.95/3902.91 & 1.557/0.983 & -0.466/-1.398\\
Si I &  3905.52         &       1.909  &       -1.092   \\
Fe I &  3906.48         &       0.110  &       -2.243   \\
Fe I &  3920.26         &       0.121  &       -1.745   \\
Fe I &  3922.91         &       0.052  &       -1.649   \\
Fe I &  3927.92         &       0.110  &       -1.594   \\
Fe I &  3930.30         &       0.087  &       -1.586   \\
Ca II K & 3933.66       &       0.000  &        0.135   \\
Al I &  3944.01         &       0.000  &       -0.638   \\
Al I &  3961.52         &       0.014  &       -0.336   \\
He I &  3964.73         &      20.616  &       -1.295   \\
Ca II H & 3968.47       &       0.000  &       -0.179   \\
H$\epsilon$ & 3970.08   &      10.200  &       -0.993   \\
Fe I &  4005.24         &       1.557  &       -0.610   \\
He I &  4026.184        &      20.964  &       -2.625   \\
He I &  4026.186        &      20.964  &       -1.448   \\
He I &  4026.186        &      20.964  &       -0.701   \\
He I &  4026.197        &      20.964  &       -1.449   \\
He I &  4026.198        &      20.964  &       -0.972   \\
Mn I &  4030.75		&       0.000  &       -0.470   \\
Mn I &	4033.06		&       0.000  &       -0.618   \\
Mn I &	4034.48 	&       0.000  &	-0.811   \\
Fe I &  4045.81         &       1.485  &        0.280   \\
Fe I/Fe I &4063.59/4063.63 & 1.557/4.103 &  0.072/-0.691   \\
Fe I &  4071.74         &       1.608  &       -0.022   \\
Sr II&  4077.71         &       0.000  &        0.167   \\
Fe I &  4078.35         &       2.609  &       -1.503   \\
H$\delta$& 4101.75      &      10.200  &       -0.753   \\
Fe I &  4132.06         &       1.608  &       -0.667   \\
Fe I/Fe I &  4143.83/4143.87 & 1.557/2.858 & -0.459/-2.126 \\
Fe I &  4154.81         &       3.368  &       -0.369   \\
Fe I &  4156.80         &       2.832  &       -0.609   \\
Fe I &  4157.91         &       0.986  &       -9.428   \\
Fe I &  4181.75         &       2.832  &       -0.180   \\
Fe I &  4187.04         &       2.450  &       -0.549   \\
Fe I &  4187.80 	&       2.426  &       -0.554   \\
Fe I/Fe I &  4198.25/4198.30 &  3.368/2.400 & -0.437/-0.719  \\
Fe I &  4199.10         &       3.047  &        0.249   \\
Fe I &  4202.03         &       1.485  &       -0.708   \\
Sr II&  4215.52         &       0.000  &       -0.145   \\
Fe I &  4222.21         &       2.450  &       -0.967   \\
Ca I &  4226.73         &       0.000  &        0.244   \\
Fe II &  4233.17        &       2.583  &       -1.995   \\
Fe I &  4235.94         &       2.425  &       -0.342   \\
Fe I &  4250.79         &       1.557  &       -0.722   \\
Cr I &  4254.33         &       0.000  &       -0.114   \\
Fe I &  4260.47         &       2.399  &       -0.018   \\
Fe I &  4271.15         &       2.450  &       -0.348   \\
Fe I &  4271.76         &       1.485  &       -0.163   \\
Cr I &  4274.80         &       0.000  &       -0.230   \\
Cr I &  4289.72         &       0.000  &       -0.360   \\
Fe I/Ti II& 4294.12/4294.10 &  1.485/1.084  & -1.113/-1.108\\
Ti II&  4300.05         &       1.180  &       -0.767   \\
Fe I &  4307.90         &       1.557  &       -0.070   \\
Fe I/Fe I &  4325.74/4325.72  &0.000 1.608  & -4.815/-0.008   \\
Mn I/Mn I &  4326.14/4326.16 &  4.666/4.194 & -1.600/-1.817  \\
H$\gamma$& 4340.48      &     -10.200  &       -0.447   \\
Cr I/Fe II &4351.76/4351.77  &  1.030/2.704   &  -0.426/-2.096\\
Mg I &  4351.91         &       4.346   &        -0.838   \\
Fe I &  4375.93         &       0.000  &       -3.031   \\
Fe I &  4383.54         &       1.485  &        0.200   \\
He I &  4387.93         &      21.218  &       -0.883   \\
Fe I &  4404.75         &       1.557  &       -0.142   \\
Fe I &  4415.12         &       1.608  &       -0.613   \\
Sr II&  4417.50         &       1.658  &       -1.676   \\
Cr I &  4424.27         &       3.011  &       -0.365   \\
Fe I/Fe I &  4427.30/4427.31 &  3.654/0.052 & -1.302/-2.907  \\
Fe I &  4447.72         &       2.223  &       -1.342   \\
Ca I &  4454.78         &       1.899  &       0.258    \\
Ca I &  4455.89         &       1.899  &       -0.526   \\
Fe I &  4459.12         &       2.176  &       -1.279   \\
Fe I &  4461.65         &       0.087  &       -3.210   \\
Fe I &  4466.55         &       2.832  &       -0.600   \\
He I &  4471.470        &      20.964  &       -2.203   \\
He I &  4471.4741       &      20.964  &       -1.026   \\
He I &  4471.4743       &      20.964  &       -0.278   \\
Mg II?& 4481.13         &       8.864  &       0.730    \\
Fe I &  4482.17         &       0.110  &       -3.500   \\
Ti II&  4501.27         &       1.116  &       -0.748   \\
Fe I &  4528.61         &       2.176  &       -0.822   \\
Fe II&  4549.47         &       2.828  &       -1.748   \\
Fe II&  4583.84         &       2.807  &       -2.019   \\
He II&  4685.70         &      48.375  &        1.181   \\
Fe I &  4701.05         &       3.686  &       -1.961   \\
Sr I &  4784.32         &       1.798  &       -0.510   \\
H$\beta$ &4861.34       &      10.200  &       -0.020   \\
Fe I &  4890.75         &       2.876  &       -0.424   \\
Fe I &  4891.49         &       2.851  &       -0.138   \\
Fe I &  4920.50         &       2.833  &        0.058   \\
He I &  4921.93         &      21.218  &       -0.435   \\
Fe II & 4923.93         &       2.891  &       -1.319   \\
Fe I &  4924.77         &       2.279  &       -2.222   \\
Fe I &  4942.95         &       0.052  &       -8.595   \\
Fe I &  4957.60         &       2.808  &        0.160   \\
Ti I &  5014.19         &       0.000  &       -1.222   \\
He I &  5015.68         &      20.616  &       -0.820   \\
Ti I &  5016.16         &       0.848  &       -0.574   \\
Fe II & 5018.44         &       2.891  &       -1.213   \\
Mg I &  5167.32         &       2.709  &       -0.856   \\
Fe I &  5167.49         &       1.485  &       -1.251   \\
Fe II & 5169.03         &       2.891  &       -0.871   \\
Mg I &  5172.68         &       2.712  &       -0.380   \\
Mg I &  5183.60         &       2.717  &       -0.158   \\
Ti I/Ti II & 5188.65/5188.68  &  2.268/1.582 & -2.930/1.210 \\
Fe I &  5204.58 	&	0.087  &	-4.332  \\
Cr I &  5208.42         &       0.941  &       0.158    \\
Ti I &  5210.39         &       0.048  &       -0.884   \\
Fe I &  5227.19		& 	1.557  &	-0.969	\\
Fe I &  5269.54         &       0.859  &       -1.322   \\
Ca I &  5270.27         &       2.526  &        0.018   \\
Fe I &  5270.36         &       1.608  &       -1.505   \\
Fe I &  5328.04         &       0.915  &       -1.465   \\
Fe I &  5328.53 	&	1.557	&	-1.653	\\
Fe I/Fe I & 5371.43/5371.49  & 4.435/0.958  &  -1.219/-1.644 \\
Fe I &  5397.13         &       0.915  &       -1.992   \\
Fe I &  5405.77         &       0.990  &       -1.844   \\
Cr I &  5409.77         &       1.030  &       -0.720   \\
Fe I &  5429.70         &       0.958  &       -1.879   \\
Fe I &  5434.52		&	1.011  &       -2.122   \\
Fe I &  5446.92         &       0.990  &       -1.928   \\
Fe I &  5455.61         &       1.011  &       -1.754   \\
Mg I &  5528.42         &       4.346  &       -0.620   \\
Fe I &  5620.49         &       4.154  &       -1.789   \\
Fe I &  5871.30         &	4.154  &       -1.991	\\
He I &  5875.60         &      20.964  &       -1.516   \\
He I &  5875.614        &      20.964  &       -0.341   \\
He I &  5875.615        &      20.964  &        0.408   \\
He I &  5875.63         &      20.964  &       -0.340   \\
Na I &  5895.92         &       0.000  &       -0.191   \\
Fe I &  6230.73         &       2.559  &       -1.281   \\
Fe I &  6462.73         &       2.453  &       -2.596   \\
He I &  6678.15         &      21.218  &        0.329   \\
Fe I &  6855.71		&       4.607  &       -1.820	\\
He I &  7065.18         &      20.964  &       -0.461   \\
He I &  7065.22         &      20.964  &       -0.682   \\
He I &  7065.71         &      20.964  &       -1.160   \\
K I  &  7664.91         &       0.000  &        0.135   \\
K I  &  7698.97         &       0.000  &       -0.168   \\
Ca I &  7707.79		&	5.578  &       -2.723   \\
Ca I &  8541.21		&	4.533  &       -1.986	\\
Ca II & 8662.14         &       1.693  &       -0.623   \\
\enddata
\tablenotetext{1}{Where blends occur and one line can not easily be ruled
out, we include these lines together on a single row.}
\end{deluxetable}

\begin{deluxetable}{lcccc}
\tabletypesize{\scriptsize}
\tablecaption{Balmer Decrement}
\tablewidth{0pt}
\tablehead{\colhead{Line} &
\colhead{BD$_{F1}$} &\colhead{FWHM$_{F1}$ (\AA)}& \colhead{BD$_{F2}$} & \colhead{FWHM$_{F2}$ (\AA)} }
\startdata
H$\beta$ & 1.19& 0.911$\pm$0.05 &1.53   & 0.690$\pm$0.05\\
H$\gamma$ & 1.00& 0.765$\pm$0.05  &1.00   & 0.510$\pm$0.1\\
H$\delta$& 0.65& 0.644$\pm$0.1  &0.45   & 0.446$\pm$0.1\\
H$\epsilon$\tablenotemark{1}& 0.34      & 0.653$\pm$0.1 &0.33& 0.415$\pm$0.2\\
H8& 0.12        & 0.794$\pm$0.2  &0.11& 0.508$\pm$0.2\\
H9\tablenotemark{2}&0.11& 0.851$\pm$0.2 &0.10& 0.400$\pm$0.3\\
\enddata
\tablenotetext{1}{Blended with Ca II H line.}
\tablenotetext{2}{Blended with several Fe I lines.}
\end{deluxetable}
 
\end{document}